\documentclass[12pt]{article}
\usepackage{amsmath}
\usepackage{graphicx}
\usepackage{amsfonts}
\usepackage{amssymb}

\begin{document}

\title{\vskip-3cm {\baselineskip14pt \centerline{\normalsize MZ-TH/00--02 \hfill}
\centerline{\normalsize\hfill} \centerline{\normalsize January 2000
\hfill}} \vskip2cm $\Delta I=\frac{1}{2}~$Enhancement and the
Glashow-Schnitzer-Weinberg Sum Rule }
\author{N. F. Nasrallah$^{1\ }$and K.~Schilcher$^{2}$\\{\normalsize $^{1}$ Faculty od Science, Lebanese University,}\\{\normalsize \ P.O. Box 826, Tripoli,Lebanon}\\[0.7ex] {\normalsize {$^{2}$ Institut f\"{u}r Physik,
Johannes-Gutenberg-Universit\"{a}t,}}\\{\normalsize {Staudinger Weg 7, D-55099 Mainz, Germany} }\\[3ex]}
\date{}
\maketitle
\begin{abstract}
In 1967 Glashow, Schnitzer and Weinberg derived a sum rule in the soft pion
and soft kaon limit relating the $\Delta I=\frac{1}{2}$ non-leptonic
$K\rightarrow2\pi$ amplitude to integrals over strange and non-strange
spectral functions. Using the recent ALEPH data from $\tau$-decay, we show
that the sum rule, slightly modified to reduce contributions near the cut,
yields the correct magnitude decay amplitude corresponding to the $\Delta
I=\frac{1}{2}$ rule.
\end{abstract}

%
%
%
%
%

%
%
%
%
%
\newpage

The $\Delta I=\frac{1}{2}$ rule for kaons has been a challenge to
theoreticians for more than four decades, see \cite{de Rafael} for a review.
\ The current-current weak non-leptonic Hamiltonian of the Standard Model
leads naively to the expectation of roughly equal $\Delta I=1/2$ and $\Delta
I=3/2$ amplitudes. \ Experimentally, however, the $\Delta I=1/2$ amplitudes
are enhanced by about a factor 20. QCD corrections may be computed with the
help of the operator product expansion yielding a $\ \Delta I=1/2$
enhancement, but its magnitude turns out too small. The situation may be
improved by the use of chiral perturbation theory combined with model input,
see \cite{prades} for a recent calculation.

In this note we will return to the roots and reanalyse an old current algebra
calculation of the $K_{s}^{0}\rightarrow2\pi$ matrix element by Glashow,
Schnitzer and Weinberg (GSW) \cite{GSW}. In this classic paper the $K_{s}%
^{0}\rightarrow2\pi$ amplitude is related to integrals over spectral functions
which were in turn evaluated by using the crude approximation of saturation by
narrow resonances. As the integrals involve differences of large numbers, it
is not astonishing that these authors do not find the right answer. Using the
$K_{s}^{0}\rightarrow2\pi$ life time as an imput they end up with a prediction
of 8 GeV for the W mass. Since precise data on the relevant spectral functions
have recently been obtained by the ALEPH collaboration \cite{aleph} it seemed
prudent to us to reanalyse the GSW formula. It should be pointed out, however,
that the ALEPH data do not saturate chiral sum rules even at s as large as
$3GeV^{2}$ if substituted directly. In fact, it was shown \cite{domschi} that
if modified chiral sum rules, based on linear combinations of spectral
function that vanish at the end of the integration region, are used the
saturation is quite spectacular.

We will repeat briefly the basic steps of the GSW calculation. In the language
of the standard model, the non-leptonic $\Delta S=1$ weak Hamiltonian is given by%

\begin{equation}
H_{W}(0)=f^{2}\int dxD_{\mu\nu}(x,M_{W}^{2})T\{j_{\mu}^{ud}(x)j_{\upsilon
}^{su}(0)+h.c.\} \label{HW}%
\end{equation}
with $f^{2}\equiv\frac{G_{F}}{\sqrt{2}}M_{W}^{2}V_{ud}V_{us}^{\ast}$ where
$M_{W}$ is the W boson mass, $D_{\mu\nu}(x,M_{W}^{2})$ its propagator, $G_{F}$
the Fermi coupling constant $(1.166\times10^{-5}GeV^{-2})$, $V_{ud}$,
$V_{us}^{\ast}$ are matrix elements of the CKM matrix \ and%

\begin{equation}
j_{\mu}^{ud}=\overline{u}\gamma_{\mu}(1-\gamma_{5})d,\,\;\;\;j_{\mu}%
^{su}=\overline{s}\gamma_{\mu}(1-\gamma_{5})u.\label{currents}%
\end{equation}
are the strangeness conserving and strangeness changing weak currents,
respectively. We consider the decay $K_{s}^{0}\rightarrow\pi^{+}\pi^{-}$ which
is described by the matrix element
\begin{equation}
\frak{M}=\left\langle \pi^{+}\pi^{-}\left|  H_{W}\right|  K_{s}^{0}%
\right\rangle \;.\label{matrixelement}%
\end{equation}

Using the conventional ''soft pion'' technique and the $SU(2)\times SU(2)$
commutation relations yields
\begin{equation}
\frak{M}=\frac{1}{4f_{\pi}^{2}}\left\langle 0\left|  H_{W}\right|  K_{s}%
^{0}\right\rangle \label{PCAC}%
\end{equation}
It should be remembered that the soft pion approach (or $SU(2)\times SU(2)$
chiral perturbation theory) gives all non-leptonic K-decay rates in terms of
$\frak{M}$, and in accord with the $\Delta I=\frac{1}{2}$ rule.

The next step in the derivation is slightly controvertial, namely, GSW
evaluate the matrix element $\left\langle 0\left|  H_{W}\right|  K_{s}%
^{0}\right\rangle $ in the soft kaon limit using the $SU(3)\times SU(3)$
algebra of currents. This step is not equivalent to evaluating the original
matrix element in the $SU(3)\times SU(3)$ chiral limit where the pions and the
kaon would have to be treated on the same footing from the beginning and where
the amplitude would vanish. Substituting the spectral representation for the
vector and axial vector correlators and exchanging the order of integration
GSW obtain finally
\begin{equation}
\frak{M}=\frac{3f^{2}}{64\pi^{2}f_{\pi}^{2}f_{K}}%
{\displaystyle\int}
ds\,s^{2}\frac{1}{2\pi^{2}}\left[  (v(s)+a(s))^{ud}-(v(s)+a(s))^{us}\right]
\left\{  \frac{\ln(s/M_{W}^{2})}{1-s/M_{W}^{2}}\right\}  \label{specrep}%
\end{equation}
where $v(s)$ and $a(s)$ are the vector and axial vector spectral functions
normalized according to $v(s)=a(s)=\frac{1}{2}(1+\frac{\alpha_{s}}{\pi}+...)$
in perurbative QCD. In the derivation of Eq.\ref{specrep} the first and second
Weinberg sum rules, which are valid in QCD, have been used.

To evaluate the integral in Eq.\ref{specrep}, which extends from threshold to
$\infty$, we split the integration range into two two parts. From threshold to
$3GeV^{2}$ we substitute ALEPH data, and from $3GeV^{2}$ to $\infty$ we use
QCD \cite{chetyrkin}
\begin{equation}
\left[  (v(s)+a(s))^{ud}-(v(s)+a(s))^{us}\right]  _{QCD}=\frac{4352}%
{727}\alpha_{s}[\alpha_{s}\left\langle \overline{u}u\right\rangle
(\left\langle \overline{u}u\right\rangle -\left\langle \overline
{s}s\right\rangle )]\frac{1}{s^{3}}\;\label{specqcd}%
\end{equation}
where facrtorization has been used for the four-quark condensate and
$\left\langle \overline{u}u\right\rangle =\left\langle \overline
{d}d\right\rangle $ was put. As the QCD part can only serve as an order of
magnitude estimate, we should make sure that the integral is saturated by the
low energy contribution accessible to experiment. Unfortunately, we know from
a detailed study \ \cite{domschi} of Weinberg's and other chiral sum rules
that this is not the case. If, however, modified sum rules build up of
suitable linear combinations of spectral function sum rules are used which
involve integrands that vanish at the end of the given finite integration
range, precocious saturation is observed to a surprising extend. This result
is understandable from the point of view of global duality in QCD. Analyticity
\ properties of the two-point function relate the integral over a spectral
function, multiplied by polynomials, to an integral over a circle in the
complex plane. Perurbative QCD is expected to break down close to the
time-like real axis, so it is desirable to choose the polynomial so as to
vanish at the end of the region of integration over the cut.

In the present application we modify the integral Eq.\ref{specrep} by adding a
term that vanishes by Weinberg's sum rule,
\begin{align}
\frak{M}  &  =\frac{3f^{2}}{64\pi^{2}f_{\pi}^{2}f_{K}}[%
{\displaystyle\int^{R}}
+\int_{R}^{\infty}]ds\,s^{2}\frac{1}{2\pi^{2}}\left[  (v(s)+a(s))^{ud}%
-(v(s)+a(s))^{us}\right] \label{modifiedsr}\nonumber \\
&  \times\left\{  \frac{\ln(s/M_{W}^{2})}{1-s/M_{W}^{2}}-\frac{R}{s}\frac
{\ln(R/M_{W}^{2})}{1-R/M_{W}^{2}}\right\}
\end{align}

If the integral is saturated precociously the integral from R to $\infty$
should be negligible. We will assume this to be the case for the time being.
In Fig. 1 we plot the result of the first integral over the experimental
spectral functions as a function of the upper limit of integration R. As the
individual contributions of the non-strange and strange spectral functions are
large, we are faced with the difference of two large numbers, and it is the
more amazing that the integral appears saturated for $R\geq1.8GeV^{2}$. The
small oscillations are typical and will level out for larger R. If the high
energy QCD contribution is neglected our prediction is therefore in agreement
with the experimental amplitude, $\left|  \frak{M}\right|  _{\exp}%
=7.78\times10^{-7}m_{K}$ , to within the errors expected from the soft meson
approximation.%
\begin{figure}
[ptb]
\begin{center}
\includegraphics[
height=2.4448in,
width=3.3883in
]%
{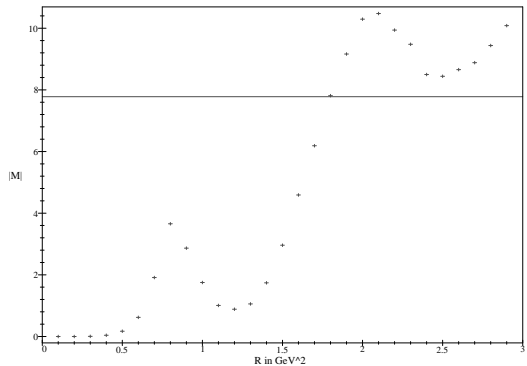}%
\caption{The decay matrix element calculated from the modified sum rule (Equ.
7) as a function of the upper limit of integration R. The solid line is the
experimental value}%
\end{center}
\end{figure}

To show that the integral from R to $\infty$ is indeed small we use the
estimates
\begin{equation}
\alpha_{s}\left\langle \overline{u}u\right\rangle ^{2}=1.9\times10^{-4}%
GeV^{6},\;\;\left\langle \overline{s}s\right\rangle =0.5\times\left\langle
\overline{u}u\right\rangle \label{condensates}%
\end{equation}
to find that the high energy QCD contribution is less than 1 \% \ for the
lower limit of integration R between 2 to 3 $GeV^{2}$, in a direction bringing
the prediction closer to the experimental value.

In Fig. 2 we plot the same integral using the original sum rule
Eq.\ref{specrep}. It is seen that the sum rule converges poorly, but the
result is still consistent with the experimental value of the decay amplitude.
This is in agreement with the analysis \cite{domschi} of the non-strange
Weinberg sum rules where also only suitable linear combinations of sum rules
that vanish at the radius R converge precociously.%
\begin{figure}
[ptb]
\begin{center}
\includegraphics[
height=2.7034in,
width=3.7472in
]%
{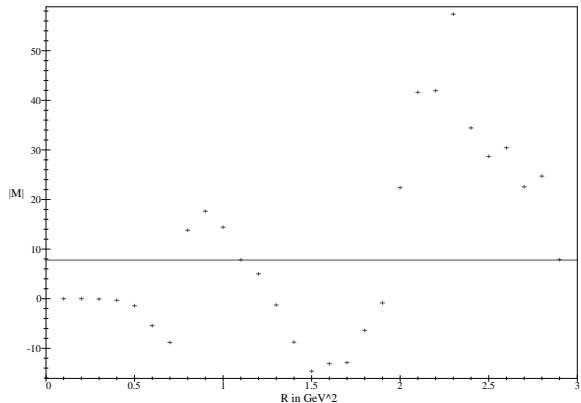}%
\caption{The decay matrix element (in units of $10^{-7}m_{K})$ as calculated
from the GSW sum rule (Equ. 5) as a function of the upper limit of integration
R. The solid line is the experimental value}%
\end{center}
\end{figure}

In Fig.1 and Fig.2 all experimental errors on the spectral functions are
suppressed. This is because the statistical part of the errors is washed out
completely upon integration and, hopefully, a great part of the (unknown)
systematic errors cancel in the differences of spectral functions. We think
that this attitude is justified a posteriori by the results plotted in the
figures which show the expected shape. We also made an estimate of the small
error made by neglecting the contribution of the charm quark.

Using the oscillation of the integral with \ $s\leq R$ \ in
\ Eq.\ref{modifiedsr} as an estimate of the total error, we arrive at a
prediction for the $K_{s}^{0}\rightarrow\pi^{+}\pi^{-}$ matrix element
\begin{equation}
\left|  \frak{M}\right|  =(9.3\pm.9)\times10^{-7}\label{ME}%
\end{equation}
as compared to the experimental value
\begin{equation}
\left|  \frak{M}\right|  _{\exp}=7.78\times10^{-7}\label{MEexp}%
\end{equation}

It is remarkable that the simple formula of GSW combined with modern spectral
function data should lead to the right prediction of the $\Delta I=\frac{1}%
{2}$ enhancement. The central value prediction is high by 15 \%, i.e. of the
order of magnitude expected from errors of the soft kaon approximation and of
the neglect of the high energy tail. Had the modern data been available to GSW
they would have predicted in 1967 the existence of a charged intermediate
vector boson of a mass of about 90 GeV (instead of the 8 GeV following from
their assumption of narrow resonance dominance). To quote GSW '' we would lose
most of our scruples about this calculation if such an intermediate boson is found''.

\end{document}